\documentclass[aps, prl, reprint, superscriptaddress]{revtex4-2}

\usepackage{graphicx, color}
\usepackage{amsmath}
\usepackage{mathtools}
\usepackage{subsupscripts,epstopdf}
\usepackage{braket}
\usepackage{graphicx}
\usepackage{dcolumn}
\usepackage{bm}
\usepackage{hyperref} 
\hypersetup{
    colorlinks=true,
    linkcolor=blue,
    citecolor=blue,
    filecolor=magenta,      
    urlcolor=blue,
    }
\usepackage[normalem]{ulem}
\begin{document}

\preprint{APS/123-QED}

\title{Simultaneous imaging of vibrational, rotational, and electronic wave packet dynamics in a triatomic molecule}

\author{Huynh Van Sa Lam}
\email[Email: ] {huynhlam@ksu.edu}
\affiliation{James R. Macdonald Laboratory, Kansas State University, Manhattan, KS 66506, USA}
\author{Van-Hung Hoang}
\affiliation{James R. Macdonald Laboratory, Kansas State University, Manhattan, KS 66506, USA}
\author{Anbu Selvam Venkatachalam}
\affiliation{James R. Macdonald Laboratory, Kansas State University, Manhattan, KS 66506, USA}
\author{Surjendu Bhattacharyya}
\affiliation{James R. Macdonald Laboratory, Kansas State University, Manhattan, KS 66506, USA}
\author{Keyu Chen}
\affiliation{James R. Macdonald Laboratory, Kansas State University, Manhattan, KS 66506, USA}
\author{Sina Jacob}
\affiliation{Institut f\"{u}r Kernphysik, Goethe-Universit\"{a}t Frankfurt, 60438 Frankfurt am Main, Germany}
\author{Sanduni Kudagama}
\affiliation{James R. Macdonald Laboratory, Kansas State University, Manhattan, KS 66506, USA}
\author{Tu Thanh Nguyen}
\affiliation{James R. Macdonald Laboratory, Kansas State University, Manhattan, KS 66506, USA}
\author{Daniel Rolles}
\affiliation{James R. Macdonald Laboratory, Kansas State University, Manhattan, KS 66506, USA}
\author{Uwe Thumm}
\email[Email: ] {thumm@ksu.edu}
\affiliation{James R. Macdonald Laboratory, Kansas State University, Manhattan, KS 66506, USA}
\author{Artem Rudenko}
\email[Email: ] {rudenko@ksu.edu}
\affiliation{James R. Macdonald Laboratory, Kansas State University, Manhattan, KS 66506, USA}
\author{Vinod Kumarappan}
\email[Email: ] {kumarappan@ksu.edu}
\affiliation{James R. Macdonald Laboratory, Kansas State University, Manhattan, KS 66506, USA}

\date{\today}

\begin{abstract}

Light-induced molecular dynamics often involve the excitation of several electronic, vibrational, and rotational states. Since the ensuing electronic and nuclear motion determines the pathways and outcomes of photoinduced reactions, our ability to monitor and understand these dynamics is crucial for molecular physics, physical chemistry, and photobiology. However, characterizing this complex motion represents a significant challenge when different degrees of freedom are strongly coupled. In this Letter, we demonstrate how the interplay between vibrational, rotational, and electronic degrees of freedom governs the evolution of molecular wave packets in the low-lying states of strong-field-ionized sulfur dioxide. Using time-resolved Coulomb explosion imaging (CEI) and quantum mechanical wave packet simulations, we directly map the bending vibrations of the molecule, show how the vibrational wave packet is influenced by molecular alignment, and elucidate the consequences of nuclear motion for the coupling between the two lowest electronic states of the cation. Our results demonstrate that multi-coincident CEI can be an efficient experimental tool for characterizing coupled electronic and nuclear motion in polyatomic molecules.

\end{abstract}

\maketitle
Recording “molecular movies” --- tracking the motion of individual atoms during ultrafast structural changes with sufficient spatiotemporal resolution --- has been a long-standing goal in molecular optical sciences \cite{Zewail2000,Ischenko2017,Ivanov_concluding_2021,zhang_ultrafast_2022}.
Remarkable progress toward this goal has been demonstrated recently, largely driven by the development of gas-phase ultrafast electron diffraction (UED) \cite{Weathersby2015,Filippetto2022} and ultrafast X-ray scattering (UXS) \cite{Minitti2015,Stankus2019,ruddock_deep_2019}, which are particularly sensitive to nuclear motion.
Together with spectroscopic methods sensitive to the electronic structure and advanced theoretical modeling, these techniques have provided critical insights into several important photochemical reactions \cite{Liu2020,Ischenko2017,Filippetto2022}.
However, despite the variety of available experimental tools, complete and unambiguous characterization of molecular dynamics remains challenging.
Central to these challenges is the ubiquitous interplay between nuclear and electronic motion, which underpins essential processes in photochemistry, photophysics, and photobiology, including energy transfer, photodissociation, and nonradiative relaxation \cite{Polli2010}. Such coupled motions are inherently complex, even for small polyatomic molecules \cite{Worner2011}. Electronic excitations reshape molecular geometries and populate a variety of vibronic levels, while nuclear-electronic couplings enable nonadiabatic transitions near conical intersections, where electronic states converge \cite{Levine2007}.

Time-resolved Coulomb explosion imaging (CEI) is another method used to create “molecular movies” \cite{Stapelfeldt1995,Dooley2003,legare2006,Matsuda2011,Ibrahim2014,Karamatskos2019,Yu2022}.
While CEI does not directly yield real-space molecular geometries, it provides excellent temporal resolution, high sensitivity to light atoms, and accessibility in university-scale laboratories. Critically, unlike UED and UXS, which project structural information onto a single coordinate (momentum transfer), coincidence-mode CEI \cite{Ullrich1997,Legare2005,Rudenko2006,Schmidt-bocking_coltrims_2021,Boll2022,Lam2024,Lam2023} generates high-dimensional data with naturally embedded correlations that enable reliable separation of reaction pathways, even for rare stochastic processes \cite{Endo2020}.
Previously, both CEI and diffraction-based techniques have been employed to directly image light-induced rotational \cite{RoscaPruna2001,Dooley2003,Lee2006,Yang2016rot,Karamatskos2019} and vibrational \cite{Rudenko2006,Fang2007,Yang2016,Glownia2016,Stankus2019} wave packets, which play a critical role in many types of molecular dynamics \cite{felker_dynamics_1985,felker_dynamics_1985-2,dantus1990,Ohmori2009,Zhang_unraveling_2019,Poullain2021,karashima2021,Rupprecht2023,karashima2024}. However, most experimental movies of molecular vibrations have been limited to diatomic molecules (such as $\mathrm{H_2^+}$ \cite{Alnaser2005, Rudenko2006}, $\mathrm{D_2^+}$ \cite{Ergler2006, Bocharova2008}, $\mathrm{N_2^+}$ \cite{Bocharova2011}, $\mathrm{O_2^+}$ \cite{De_2020, Bocharova2011}, $\mathrm{CO^+}$ \cite{De2011,Bocharova2011}, and $\mathrm{I_2}$ \cite{Fang2007, Yang2016, Glownia2016}) or the stretching of just one bond in polyatomic molecules \cite{Erattupuzha2016,rudenko_2016, Malakar2019}. Due to the fast time scale and complexity of rovibronic motions in polyatomic molecules, only a limited number of direct structural measurements of other vibrational modes has been reported \cite{Madsen2009, Christensen2014, Stankus2019}.

In this Letter, we demonstrate that laser-induced Coulomb explosion can directly image ultrafast coherent wave packets in a small polyatomic molecule and produce movies that show the motion of individual atoms even in the presence of coupled nuclear and electronic motion. We investigate sulfur dioxide (SO$_2$), a bent triatomic molecule that has been studied extensively (see, for instance, \cite{Watson1931, Clements1935, Legare2005, legare2006, leveque2013, Wilkinson2014, Mai2014, Leveque2014, spector2014, salen_complete_2015, chen_fragmentation_2023}) due to its significance in atmospheric processes and its intriguing photochemistry, including the efficient formation of neutral and ionic molecular oxygen \cite{Lin2020, Wallner2022, Rosch2023, Li2024}.  We employ time-resolved coincident CEI to create a movie of the coherent vibrational wave packet in SO$_2^+$  and show that its dynamics are significantly influenced by the rotational motion. Moreover, we find that a secondary wave packet resulting from a resonant dipole coupling between the two lowest electronic states of SO$_2^+$ traverses a conical intersection and leaves clear signatures of the nonadiabatic coupling between these two states in the ion momentum distributions.

\begin{figure}
\includegraphics[width=\columnwidth]{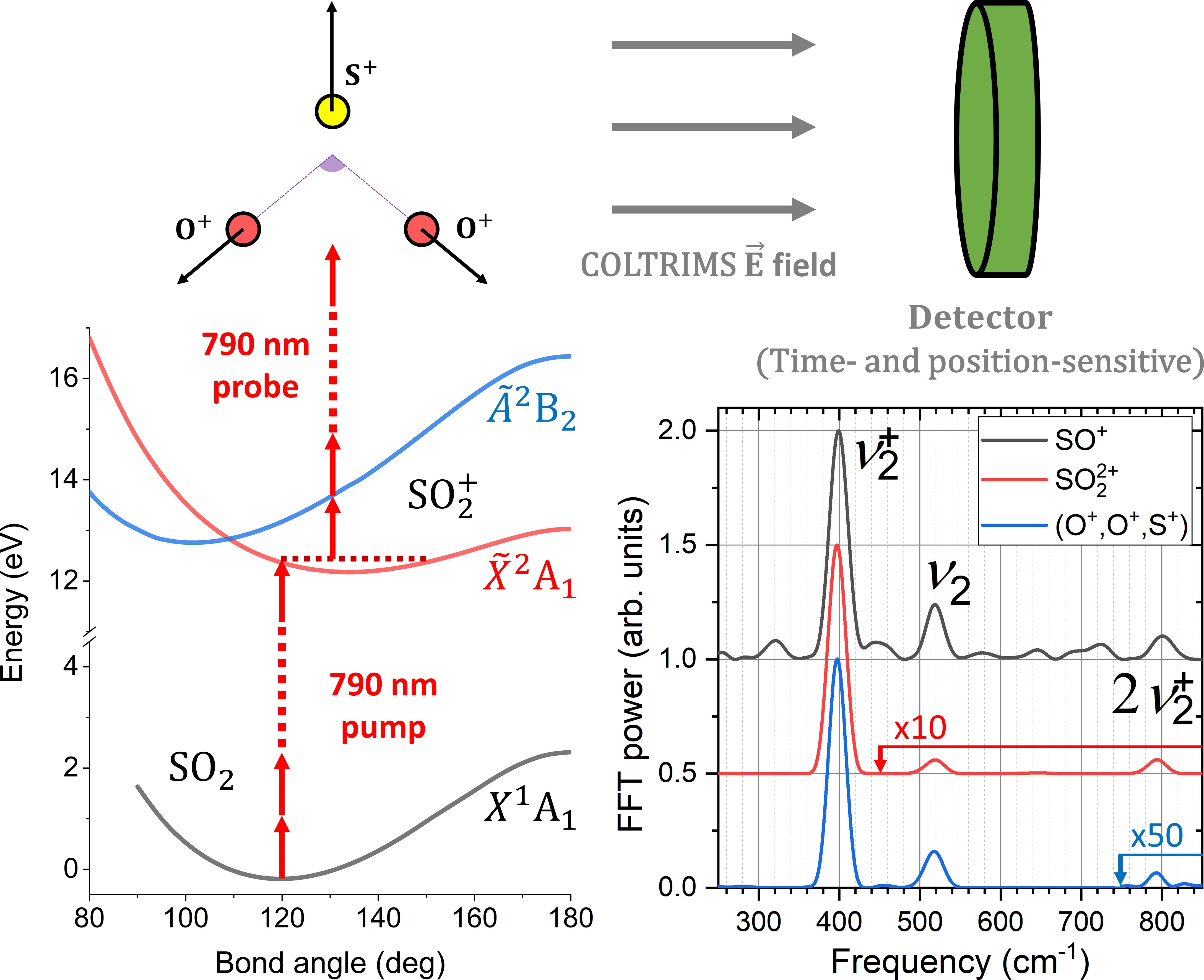}
  \caption{Pump-probe scheme and relevant potential energy curves. The bottom-right inset shows FFT spectra of the delay-dependent yield of SO$^+$, SO$_2^{2+}$, and the three-body (O$^+$, O$^+$, S$^+$) channel. For the three-body channel, only events that have the angle between two O$^+$ momentum vectors from 80$^\circ$ to 115$^\circ$ were selected.}
  \label{fig:PES_FFT}
\end{figure}

Our experiment utilized a Ti:Sapphire laser operating at 10 kHz, generating near-infrared (NIR) pulses with a central wavelength of approximately 790 nm and a pulse duration of about 28 fs. We split this output into pump and probe pulses, with independently controlled power, and scan their relative delay using a motorized linear stage. As depicted in Fig.~\ref{fig:PES_FFT}, we use a 3.4$\times10^{14}$~W/cm$^2$ pump pulse to ionize SO$_2$ and a 8$\times10^{14}$~W/cm$^2$ probe pulse to further ionize and dissociate (i.e., Coulomb explode) it.
The ions are then detected in coincidence using a COLTRIMS apparatus \cite{Maharjan2007}. The momenta of these ions, determined from their measured time of flight and impact position, are used to deduce information about the induced wave packets.

Previous studies \cite{Eland1968, Wang1987, Holland1994, Mo2004} have reported that the bending mode progression dominates the SO$_2^+$~($\tilde{X}^2A_1$) $\leftarrow$ SO$_2$~($X^1A_1$) photoelectron spectra because the equilibrium geometries in these electronic ground states have almost identical bond lengths but different bond angles \cite{Eland1968, Zhang2007}. This leads to a strong excitation into the bending mode ($\nu_2^+$) and very weak excitation into the symmetric ($\nu_1^+)$ and asymmetric ($\nu_3^+$) stretch modes. The first seven bending-mode energies are approximately equally spaced with a nearest-neighbor separation of $\approx400$ cm$^{-1}$. In our experiment, which has a 17~cm$^{-1}$ frequency resolution, we see oscillations at this frequency (and its first overtone at $\approx800$ cm$^{-1}$) in different observables for various ionization and fragmentation channels. This indicates that these channels involve the population of the SO$_2^+$ ground state by the pump pulse and subsequent ionization and/or fragmentation by the probe. For example, the fast Fourier transform (FFT) spectra of the delay-dependent yields of SO$^+$, SO$_2^{2+}$ and $\mathrm{(O^+,O^+,S^+)}$ channels show a strong peak at the bending frequency and a much smaller one at the overtone (Fig.~\ref{fig:PES_FFT}). They also show a weak but clear signature of the bending vibration in the ground state of neutral SO$_2$ (at 520~cm$^{-1}$) \cite{Polo1954}. The latter could be induced by Raman excitation (analogous to bond softening), or by geometry-dependent ionization (so-called ``Lochfrass" process) \cite{Goll2006, Ergler2006, Fang2008, rudenko_2016, Wei2017}. The determination of the relative importance of these mechanisms requires a detailed analysis of the wave packet phase and will be discussed elsewhere. Here, we focus on the three-body channel to directly visualize the bending vibrations of the cation via delay-dependent kinetic energy (KE) spectra and angle correlations between the fragment ions.

\begin{figure*}
\includegraphics[width=\textwidth]{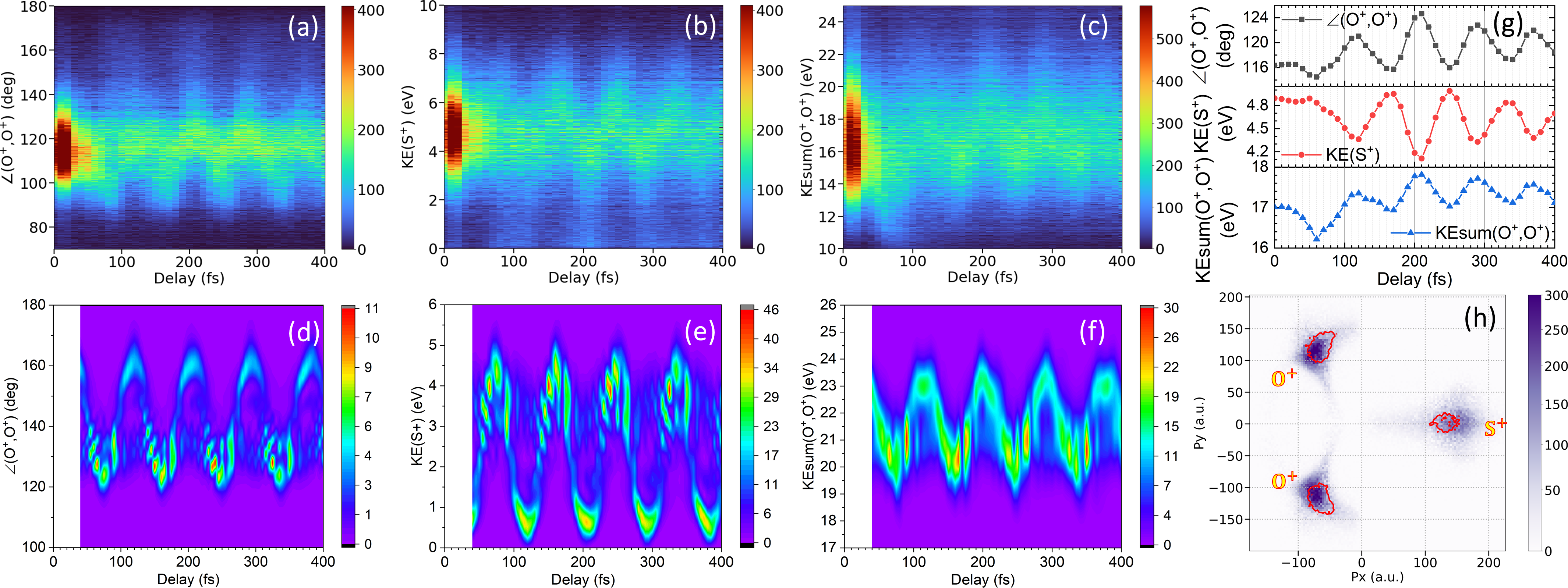}
  \caption{Signatures of bending vibrations in SO$_2^+ (\tilde{X})$: Measured delay dependence of (a) angle between the two $\mathrm{O^+}$ momenta, (b)  KE of $\mathrm{S^+}$, and (c) total KE of the two $\mathrm{O^+}$ fragments. Panels (d-f) show simulations of the observables in (a-c); note the different vertical scales for each pair. Panel (g) presents the mean of experimental observables in (a-c) as a function of pump-probe delay. Panel (h) shows the Newton plot at 250~fs when the bending wave packet reaches the inner turning point. The red contour (at 1/3 maximum intensity) represents the distribution at 210~fs when the bending wave packet reaches the outer turning point. The $x$ axis is defined as the bisector between the two O$^+$ momenta, and the vector difference between them defines the $xy$ plane \cite{matsuda2014}. A timed sequence of these images illustrating the motion of individual atoms is provided in the Supplemental Material (SM, which includes Refs.~\cite{Lam2021D, makhija2016, Lam2020, Wangjam2021, Lam2022}). In all 2D plots, the color bar indicates the yield. To emphasize the pump-induced effects, we subtracted the probe-only contributions in panels (a-c) and (h). No subtraction was used in subsequent analyses, and the original data is provided in Fig.~S1 of the SM. We discarded events with KER~$<$~14~eV to remove dissociating molecules.}
  \label{fig:Results}
\end{figure*}

In Fig.~\ref{fig:Results}, we present experimental data from the $\mathrm{(O^+,O^+,S^+)}$ channel illustrating the delay dependence of several key observables: (a) the angle between the momentum vectors of two O$^+$ ions [$\mathrm{\angle (O^+,O^+)}$], (b) the KE of S$^+$ fragment [KE($\mathrm{S^+}$)], and (c) the total KE of two O$^+$ fragments [KEsum($\mathrm{O^+,O^+}$)]. Simulation results are depicted in panels (d-f). The delay-dependent mean values of experimental data in (a-c) are shown in panel (g). Panel (h) features distributions of the ions at 250 fs (scatter plot) and 210 fs (contour line) when the bending wave packet is at the inner and outer turning points, respectively.

There are two distinguished features in the experimental data shown in Fig.~\ref{fig:Results}(a-c). First, all observables---$\mathrm{\angle (O^+,O^+)}$, KE($\mathrm{S^+}$), and KEsum($\mathrm{O^+,O^+}$)---show pronounced oscillations with a period of 83 fs (or 400 cm$^{-1}$), clearly indicating that we mainly image the vibrational wave packet in the ground state of SO$_2^+$ \cite{Wang1987}. Second, there is a clear correlation between these observables: $\mathrm{\angle (O^+,O^+)}$ and KEsum($\mathrm{O^+,O^+}$) oscillate in-phase with each other and out-of-phase with KE(S$^+$). This can be explained in terms of Coulomb forces between point charges (see Fig. S3 in the SM): near the equilibrium geometry, the net force on the S$^+$ (O$^+$) fragment due to the other two decreases (increases) with increasing bond angle. Moreover, $\mathrm{\angle (O^+,O^+)}$ also increases monotonically with the real-space bond angle.

We numerically model the experimental observables in Figs.~\ref{fig:Results}(a-c) by solving the time-dependent Schr\"{o}dinger equation for the coupled nuclear motion on the Born-Oppenheimer potential energy surfaces (BOS) of the SO$_2$($X$) and SO$_2^+$($\tilde{X}$) states, including all vibrational degrees of freedom (symmetric stretch, bending, and antisymmetric stretch modes)~\cite{HungPRA2024}. 
We calculate the BOS \textit{ab initio} by applying the multi-configurational self-consistent-field method as implemented in the quantum chemistry code GAMESS~\cite{GAMESS2020}, based on seven frozen inner orbitals and 12 active orbitals expanded in the correlation consistent-polarized valence triple zeta (cc-pVTZ) basis set.
We account for single ionization of SO$_2$($X$) to SO$_2^+$($\tilde{X}$) by propagating the nuclear wave functions of the neutral and cationic states subject to dipole coupling by the pump pulse, with the assumption that the photoelectron has zero kinetic energy \cite{HungPRA2024}. Within this \textit{ad hoc} dipole-coupling model, the neutral ground state is depleted (and the ionic state populated) throughout the pulse, resulting in coherent vibrational wave packets in both states \cite{Goll2006, Ergler2006}.
To account for multiple ionization in the delayed probe pulse, we project the SO$_2$ and SO$_2^+$ nuclear wave functions onto the potential energy surface of the triply-charged molecule, approximated as purely Coulombic. We then propagate the ensuing nuclear wave packet representing the S$^+$ and the two O$^+$ ions to sufficiently large internuclear distances to reach numerical convergence of their momentum expectation values. We obtain the fragment KEs by FFT of the real-space nuclear wave function to momentum space~\cite{Maia_JPB_2014}.

Our simulations shown in Figs.~\ref{fig:Results}(d-f) closely resemble the experimental data in Figs.~\ref{fig:Results}(a-c), manifesting similar periodicity, phases, and correlations between the three observables. However, each pair of experiment-simulation plots shows a mismatch in absolute values of angle or KE. The primary reason for this mismatch is that vertical projection to and propagation on the Coulomb potential, which we used in our simulation, do not accurately describe the ionization and fragmentation process caused by the probe pulse. Furthermore, as will be discussed below, the wave packet propagation on the $\tilde{A}^2B_2$ state, which extends to smaller bond angles and is not included in our simulation, also contributes to the experimental data (see also Fig. S3 in the SM).
Nevertheless, our results provide a direct intuitive picture of the strong-field-induced vibrational wave packet motion in a triatomic molecule. Modulations of this three-body channel are dominated by the ionic ground-state vibrations, with a weak contribution from bending vibrations in the ground state of neutral SO$_2$, as revealed by FFT of the delay-dependent data (Fig.~\ref{fig:PES_FFT} and SM, Fig.~S2).

Besides revealing the states involved in the vibrational wave packet, coincidence momentum imaging also allows us to visualize the correlated motion of individual atoms. To achieve this, we plot the data for all ions from the $\mathrm{(O^+,O^+,S^+)}$ channel in a 2D momentum image (Newton plot). Fig.~\ref{fig:Results}(h) features a representative image at 250 fs with a contour for data at 210 fs (the inner and outer turning points of the bending wave packet). Because the $\mathrm{\angle (O^+,O^+)}$ and KE($\mathrm{S^+}$) manifest the same behavior as the OSO bond angle and the displacement of S from the center of mass (Fig.~\ref{fig:Results} and Fig.~S3), movies comprising a timed sequence of these images (as provided in the SM) are a striking and intuitive representation of the bending vibration of the molecule. These molecular movies illustrate that during the first 400 fs, the wave packet is well localized and mimics the motion of a “ball-and-stick” model.

\begin{figure}
\includegraphics[width=\columnwidth]{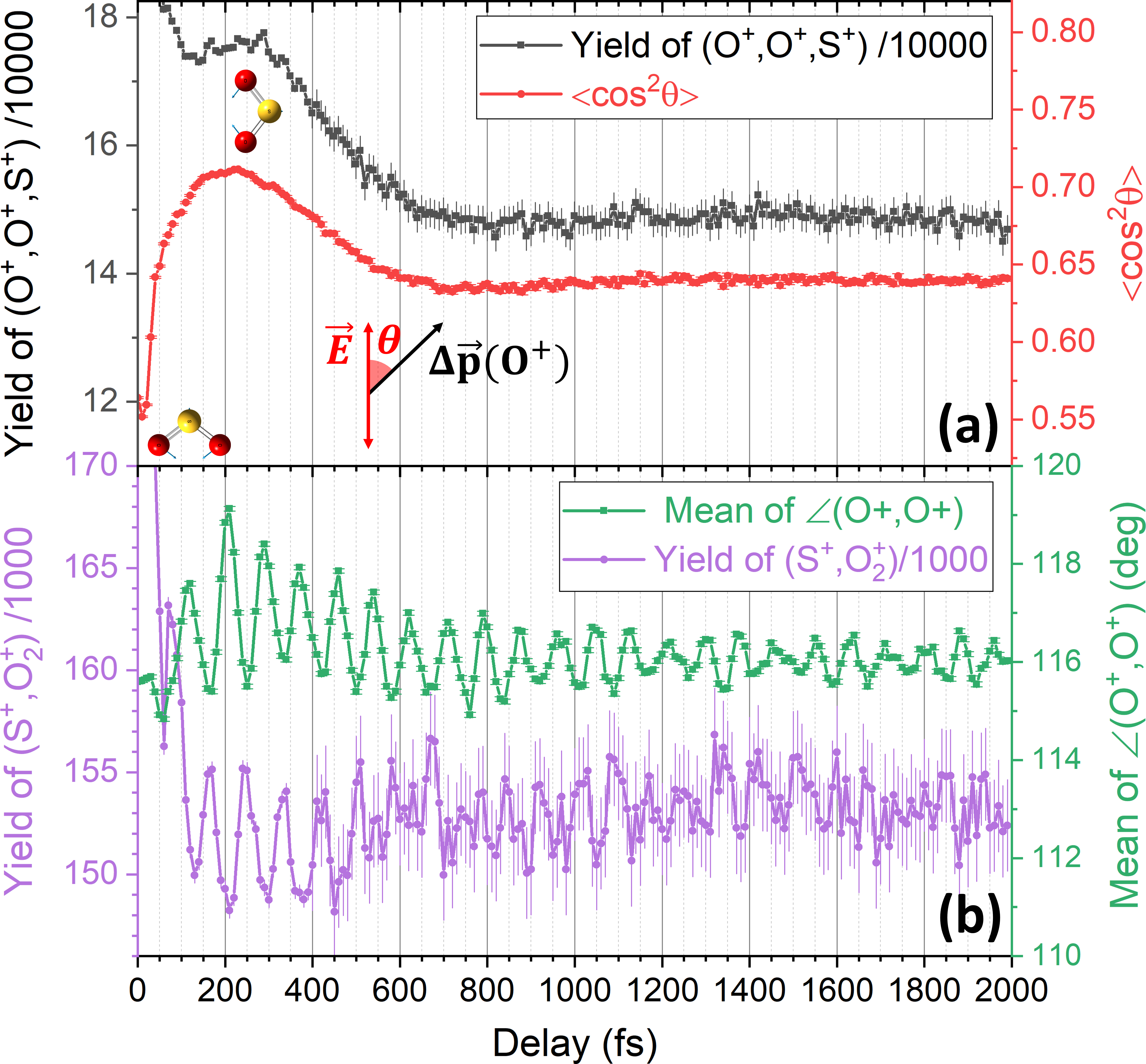}
  \caption{(a) Delay dependence of the yield (black) and $\langle \cos^2{\theta} \rangle$ (red) for the (O$^+$, O$^+$, S$^+$) channel, where $\theta$ is the angle between the laser polarization and $\Delta\vec{p}(\mathrm{O}^+)$. $\Delta\vec{p}(\mathrm{O}^+)$ is the vector difference between two O$^+$ momentum vectors, employed to represent the O-O axis. (b) The average $\mathrm{\angle (O^+,O^+)}$ in the (O$^+$, O$^+$, S$^+$) channel (green) and yield of (S$^+$,O$_2^+$) channel (purple) as functions of delays. Delays exceeding 400~fs have less statistics, resulting in larger error bars.}
  \label{fig:rot_n_asym}
\end{figure}

While bending vibrations are responsible for the most pronounced features of the observables shown in Figs.~\ref{fig:PES_FFT} and \ref{fig:Results}, the experimental data contain more information about the molecular wave packet launched by the pump pulse. One hint of other dynamics is found in the total yield of the (O$^+$, O$^+$, S$^+$) channel, which does not exhibit oscillations at the frequency of the bending vibration. This suggests that double ionization of the cation is insensitive to this mode. However, as depicted in Fig.~\ref{fig:rot_n_asym}(a), the yield initially decreases by approximately 15\% within the first few hundred femtoseconds before flattening out. To explain this behavior, we also plot $\langle \cos^2{\theta} \rangle$ for this channel in Fig.~\ref{fig:rot_n_asym}(a). Here, $\theta$ is the angle between the laser polarization and $\Delta\vec{p}(\mathrm{O}^+)=\vec{p_1}(\mathrm{O}^+)-\vec{p_2}(\mathrm{O}^+)$, representing the direction of the O-O axis. The strong correlation between these two quantities (except near the overlap region) indicates that the observed yield variation is due to the initial alignment of the pump-induced rotational wave packet. The time scale is consistent with the rotational dynamics of the two-body breakup channels shown in Fig.~S4 (SM). Although the O-O axis is the most polarizable axis of the molecule, strong-field ionization of neutral SO$_2$ is more likely to occur along the symmetry axis (see \cite{spector2014} and SM, Fig.~S5). Thus, SO$_2^+$ ions are most likely created near the peak of the pump pulse with their symmetry axes aligned with the laser polarization. They then rotate due to the angular momentum accumulated in the pump pulse and reach peak alignment (of the O-O axis) around 200~fs. After the peak, the rotational wave packet dephases and exhibits long-term incoherent alignment.

\begin{figure}
\includegraphics[width=\columnwidth]{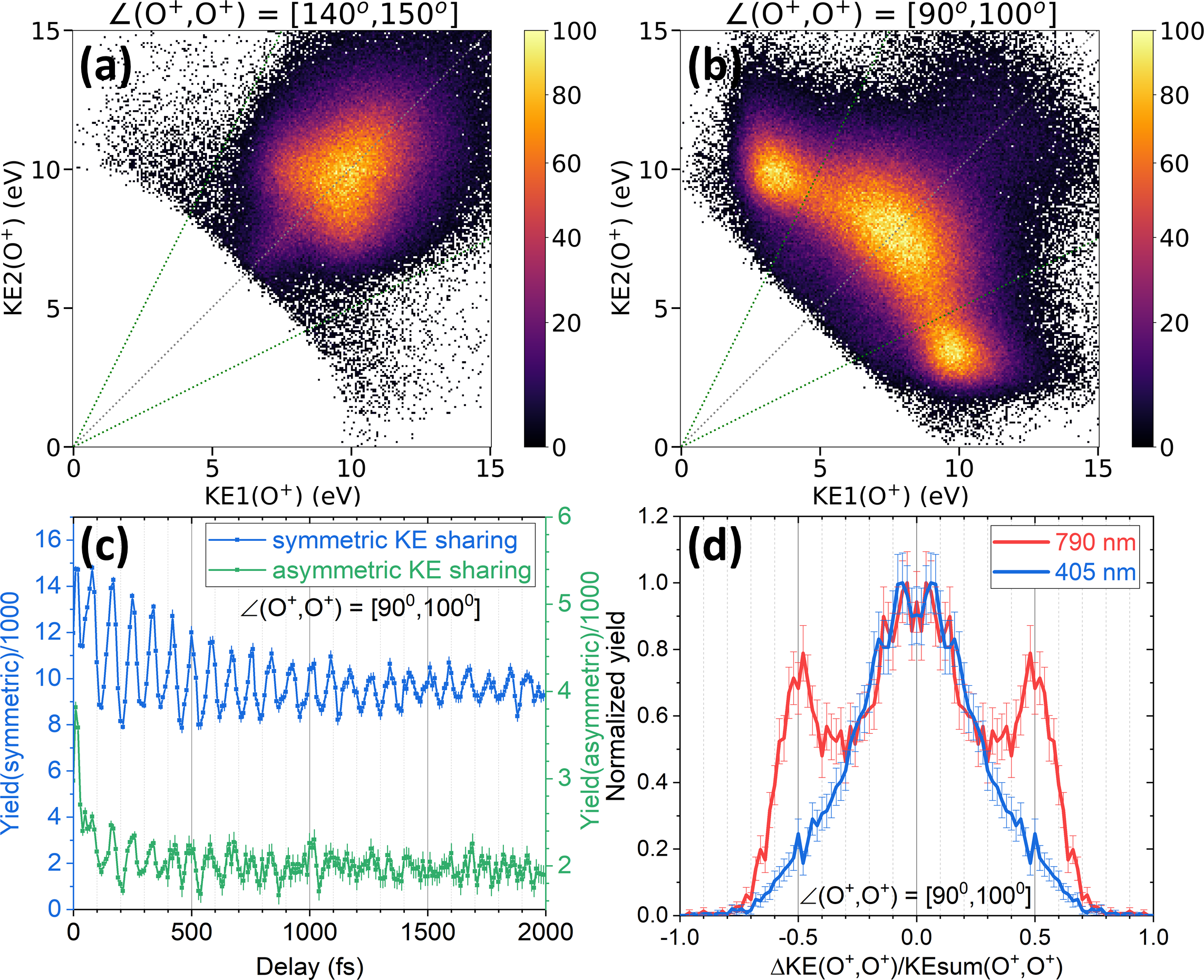}
  \caption{(a,b) KE sharing between the two O$^+$ ions for events with delay $\ge$~200~fs and large (140$^\circ$\textemdash150$^\circ$) or small (90$^\circ$\textemdash100$^\circ$) $\mathrm{\angle (O^+, O^+)}$, respectively. (c) Delay-dependent yield of events with a small $\mathrm{\angle (O^+, O^+)}$, separated by symmetric and asymmetric KE sharing between the two O$^+$ ions, defined as $\beta$=$\Delta$KE(O$^+$,O$^+$)/KEsum(O$^+$,O$^+$). Symmetric events have $\beta$~$\le$~1/3, and asymmetric events have $\beta$ $>$ 1/3. The separation is represented by the dotted green lines in (a,b). (d) Distribution of events with small $\mathrm{\angle (O^+, O^+)}$ as a function of $\beta$ following ionization by a single 790~nm or 405~nm laser pulse (obtained via second harmonic generation with a BBO crystal with an estimated intensity and pulse duration of 8$\times10^{14}$~W/cm$^2$ and 30~fs).}
  \label{fig:CI}
\end{figure}

This rovibrational wave packet motion in SO$_2^+$ has other manifestations that can be elucidated by multi-coincidence momentum imaging. One example is the formation of O$_2^+$, which can be traced by analyzing the (S$^+$,O$_2^+$) coincidence channel. The delay-dependent yield of this channel is shown in Fig.~3(b) along with the mean value of $\mathrm{\angle (O^+,O^+)}$ in the three-body channel. Clearly, the two oscillate out of phase, suggesting an intuitive picture of O$_2^+$ formation by the probe pulse: O$_2^+$ is more likely to be formed if the probe pulse finds the cation at smaller OSO angles, where the two oxygen atoms are closer to each other. The alignment and bending angle dependence of O$_2^+$ production may help improve our understanding of a potential source of abiotic oxygen in SO$_2$-rich planetary atmospheres \cite{Wallner2022}.

The rovibrational wave packet also controls the excitation of the ground-state cations to the first excited state in the probe pulse. This single-photon transition is energetically allowed for $\phi \lesssim 134^\circ$ and its dipole moment is along the O-O axis. Thus,  when ground-state cations have rotated into alignment with the probe laser polarization and bending mode vibration has pushed the nuclear probability density to smaller-then-equilibrium bond angles ($\phi < 134^\circ$), a single photon in the leading edge of the probe can resonantly transfer population from $\tilde{X} ^2A_1$ to $\tilde{A}^2B_2$ (Fig.~\ref{fig:PES_FFT}). These two states have a symmetry-allowed conical intersection (CI) at $\approx 108^\circ$ with the branching space comprised of the bending and asymmetric stretching coordinates \cite{Leveque2014}. After excitation to $\tilde{A}^2B_2$ the wave packet travels towards the CI where it is nonadiabatically coupled to the asymmetric stretch mode and returns to $\tilde{X} ^2A_1$ within 20~fs \cite{Leveque2014}. Since our probe pulse width is 28~fs, these cations can still be ionized at the peak of the probe and appear in the (O$^+$, O$^+$, S$^+$) channel. On the other hand, if the excitation occurs in the trailing edge of the pump pulse, the wave packet should dephase rapidly as it traverses the CI and not exhibit 400~cm$^{-1}$ oscillations.

In the experimental data, signatures of a wave packet in $\tilde{A}$ are seen in the KE sharing plot for the O$^+$ ions. We separate events with $\mathrm{\angle (O^+,O^+)} \in [140^\circ,150^\circ]$ and $\mathrm{\angle (O^+,O^+)} \in [90^\circ,100^\circ]$ in Fig.~\ref{fig:CI}(a,b), respectively, with the expectation that the population excited to $\tilde{A}$ can travel to smaller angles than the inner turning point of the bending wave packet in $\tilde{X}$, but not to the outer turning point, within the probe pulse. The strong component with asymmetric KE sharing seen in Fig.~\ref{fig:CI}(b) but not in Fig.~\ref{fig:CI}(a) confirms the dipole coupling to $\tilde{A}$ and nonadiabatic coupling to the asymmetric stretch mode. Fig.~\ref{fig:CI}(c) shows the delay-dependent yields of events with symmetric (blue) and asymmetric (green) KE sharing [all with small  $\mathrm{\angle (O^+,O^+)}$]. The in-phase oscillation of the two curves is consistent with population transfer by the probe to $\tilde{A}$ near the inner turning point of the ground state bending vibration. To further test this hypothesis of resonant inter-state coupling, we use a single laser pulse at 405 nm to Coulomb-explode SO$_2$ into (O$^+$, O$^+$, S$^{+}$). At this shorter wavelength, efficient dipole coupling can occur only near the outer turning point of the $\tilde{X}$ wave packet and is unlikely to occur within our 30-fs pulse. The absence of asymmetric KE sharing peaks in the 405-nm case [Fig.~\ref{fig:CI}(d)] confirms a resonant coupling near 790\,nm, corroborating our conclusion that we see signatures of mode-coupling at the CI in our pump-probe experiment.

In conclusion, our work highlights the direct sensitivity of CEI to changes in the spatial density of the nuclear wave packet propagating on multiple electronic states, demonstrating the feasibility of tracking correlated atomic motion within a molecule.
Our multidimensional coincident measurement allows us to observe and interpret the interplay between rotational, vibrational, and electronic motion, which determines the molecular wave packet evolution and experimental observables.
Notably, correlated fragment momenta reveal clear signatures of a conical intersection that couples the bending and asymmetric stretching coordinates --- which is difficult to access with other experimental techniques.
This study paves the way for tracking coupled electronic and nuclear dynamics in medium-size polyatomic molecules, where high-order correlations revealed by CEI provide unique insights into the 3D correlated motion of multiple atomic constituents.
Recent advancements have shown that CEI can image detailed 3D structures of gas-phase molecules with about ten atoms \cite{Boll2022, Lam2024, Wang2023, Jahnke2025}, bringing this goal closer to reality.

\begin{acknowledgments}
We thank Charles Fehrenbach for taking care of PULSAR laser operation. We are grateful to the technical staff of the J.R.~Macdonald Laboratory for their support. This work and the operation of the J.R.~Macdonald Laboratory are supported by the Chemical Sciences, Geosciences, and Biosciences Division, Office of Basic Energy Sciences, Office of Science, U.S. Department of Energy, Grant no.~DE-FG02-86ER13491. The PULSAR laser was provided by Grant no.~DE-FG02-09ER16115, S.B.~is supported by Grant no.~DE-SC0020276 from the same funding agency, A.S.V.~by the National Science Foundation Grant no.~PHYS-2409365. S.J.~was funded by the RISE Worldwide program of German Academic Exchange Service (DAAD) for her summer research at Kansas State University.
\end{acknowledgments}

\bibliography{2024_SO2_References.bib}

\end{document}


\preprint{APS/123-QED}

\title{Supplemental Material for\\
Imaging coupled vibrational, rotational, and electronic wave packet dynamics\\ in a triatomic molecule}

\author{Huynh Van Sa Lam}
\email[Email: ] {huynhlam@ksu.edu}
\affiliation{James R. Macdonald Laboratory, Kansas State University, Manhattan, KS 66506, USA}
\author{Van-Hung Hoang}
\affiliation{James R. Macdonald Laboratory, Kansas State University, Manhattan, KS 66506, USA}
\author{Anbu Selvam Venkatachalam}
\affiliation{James R. Macdonald Laboratory, Kansas State University, Manhattan, KS 66506, USA}
\author{Surjendu Bhattacharyya}
\affiliation{James R. Macdonald Laboratory, Kansas State University, Manhattan, KS 66506, USA}
\author{Keyu Chen}
\affiliation{James R. Macdonald Laboratory, Kansas State University, Manhattan, KS 66506, USA}
\author{Sina Jacob}
\affiliation{Institut f\"{u}r Kernphysik, Goethe-Universit\"{a}t Frankfurt, 60438 Frankfurt am Main, Germany}
\author{Sanduni Kudagama}
\affiliation{James R. Macdonald Laboratory, Kansas State University, Manhattan, KS 66506, USA}
\author{Tu Thanh Nguyen}
\affiliation{James R. Macdonald Laboratory, Kansas State University, Manhattan, KS 66506, USA}
\author{Daniel Rolles}
\affiliation{James R. Macdonald Laboratory, Kansas State University, Manhattan, KS 66506, USA}
\author{Uwe Thumm}
\affiliation{James R. Macdonald Laboratory, Kansas State University, Manhattan, KS 66506, USA}
\author{Artem Rudenko}
\email[Email: ] {rudenko@ksu.edu}
\affiliation{James R. Macdonald Laboratory, Kansas State University, Manhattan, KS 66506, USA}
\author{Vinod Kumarappan}
\email[Email: ] {vinod@phys.ksu.edu}
\affiliation{James R. Macdonald Laboratory, Kansas State University, Manhattan, KS 66506, USA}

\date{\today}

\begin{abstract}
In this Supplemental Material, we provide information that is useful to a subset of readers (who are interested in more specialized content) but is not essential to comprehend the main results of the article. The Supplemental Material includes the following figures: Original data before subtraction, additional results on rotation, simulation showing correlations between observables as a function of the bond angle, and angle-dependent strong-field ionization rate from SO$_2$ to SO$_2^+$ by a linearly polarized intense laser pulse. We also provide the molecular movies as media files, showing the correlated motion of individual atoms within the molecule.
\end{abstract}

\maketitle

\begin{figure}
\includegraphics[width=\columnwidth]{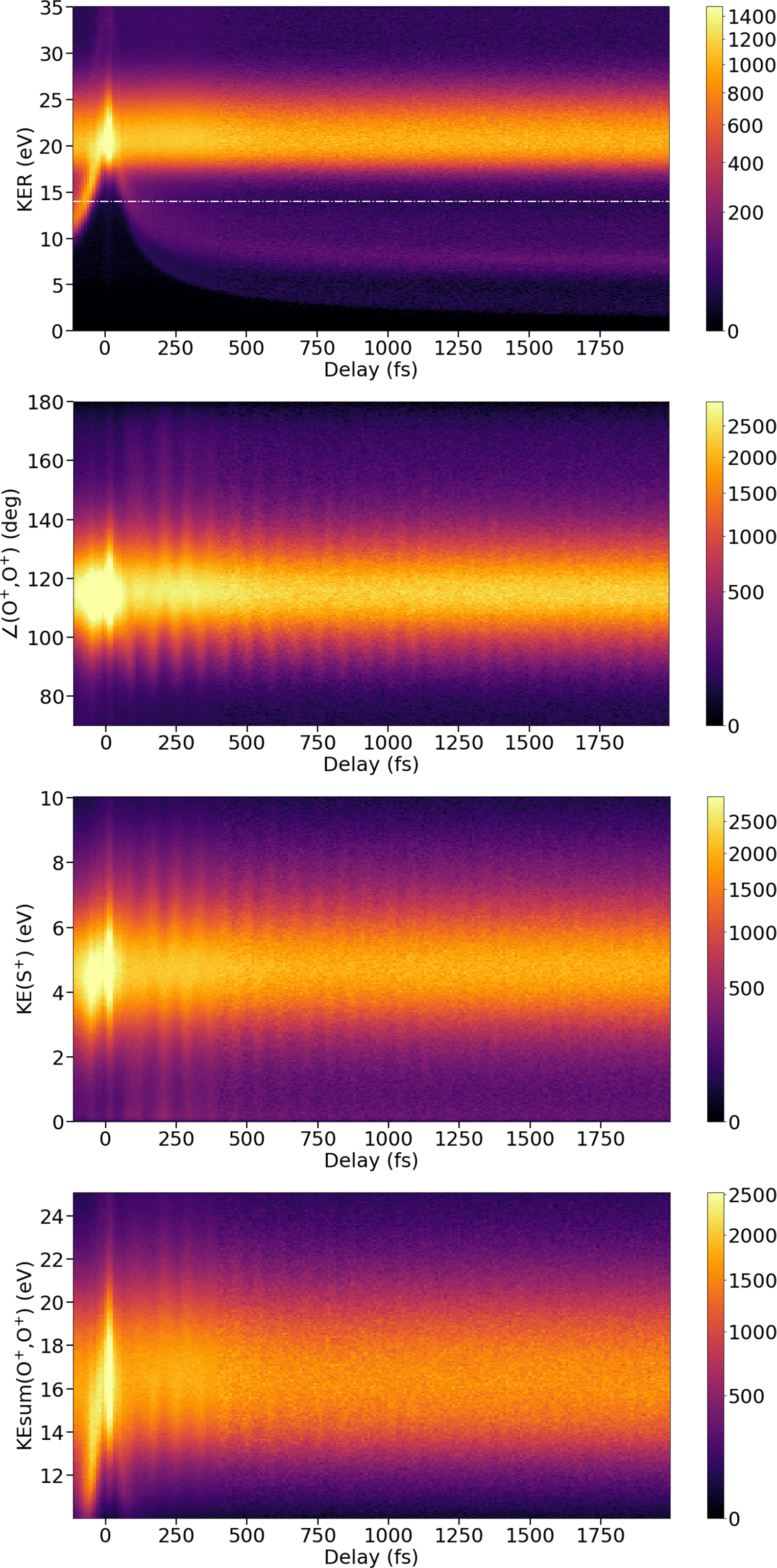}
  \caption{Original data before subtraction. From top to bottom: total kinetic energy release (KER), $\mathrm{\angle (O^+,O^+)}$, KE(S$^+$), and KEsum(O$^+$,O$^+$). The horizontal dashed line in the KER spectrum is at 14\,eV. In the main paper, we analyze events with KER~$>$~14\,eV. It is worth noting that since KE($\mathrm{S^+}$) and KEsum($\mathrm{O^+,O^+}$) are out of phase, the total kinetic energy release (KER, which is the sum of these two) does not oscillate as a function of delay. The total yield of the three-body coincidence channel also remains relatively flat. These two observables do not show the signature of bending vibration.}
  \label{fig:data}
\end{figure}
\begin{figure}
\includegraphics[width=\columnwidth]{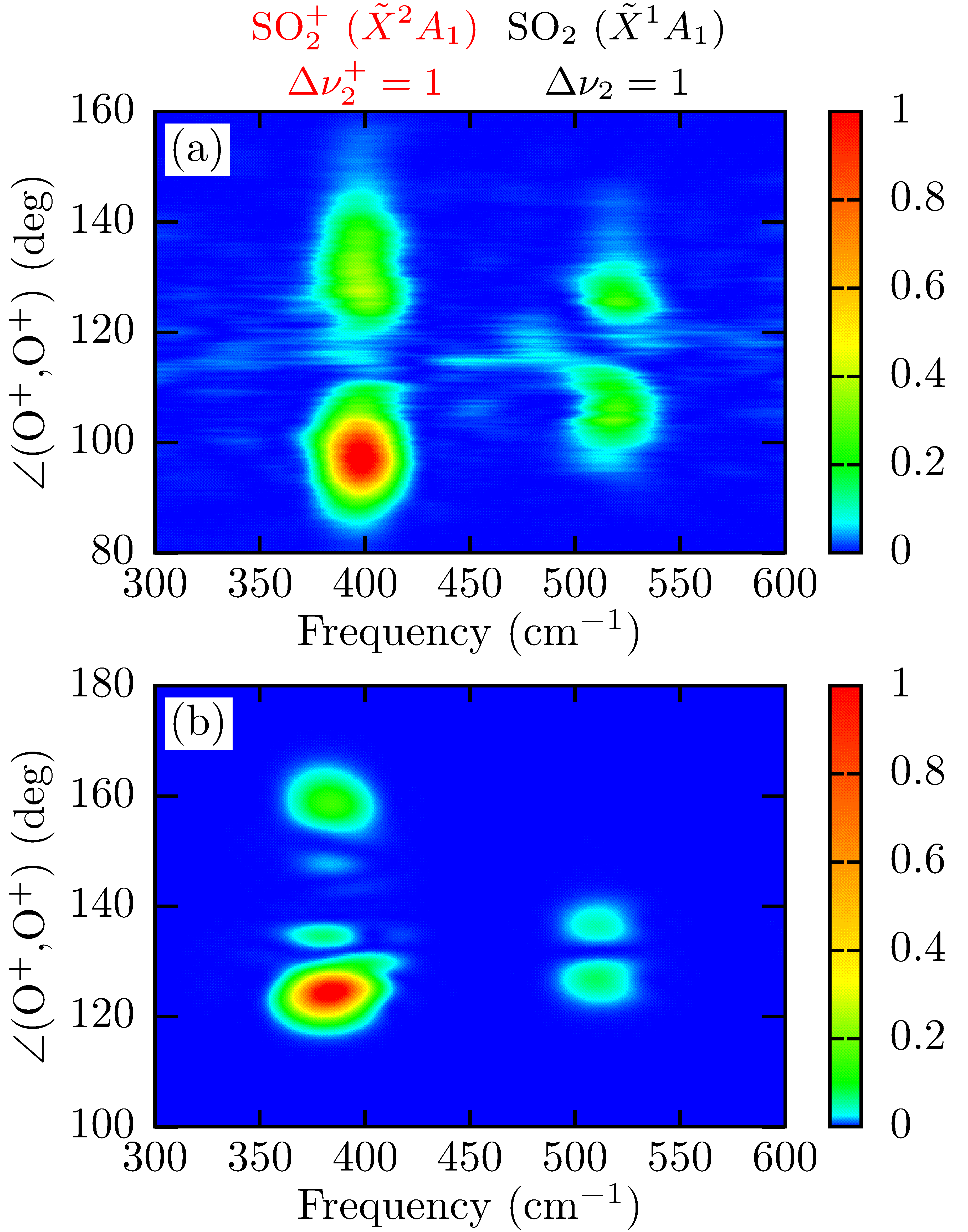}
  \caption{FFT spectra as a function of $\angle$(O$^+$,O$^+$) from experimental data (top) and theoretical simulation (bottom). Although clearly dominated by the ionic ground-state vibrations, the FFT spectra also manifest a weaker but clear signature of the ground-state bending vibrations in the neutral SO$_2$ molecule.} 
  \label{fig:fft}
\end{figure}
\begin{figure}
\includegraphics[width=\columnwidth]{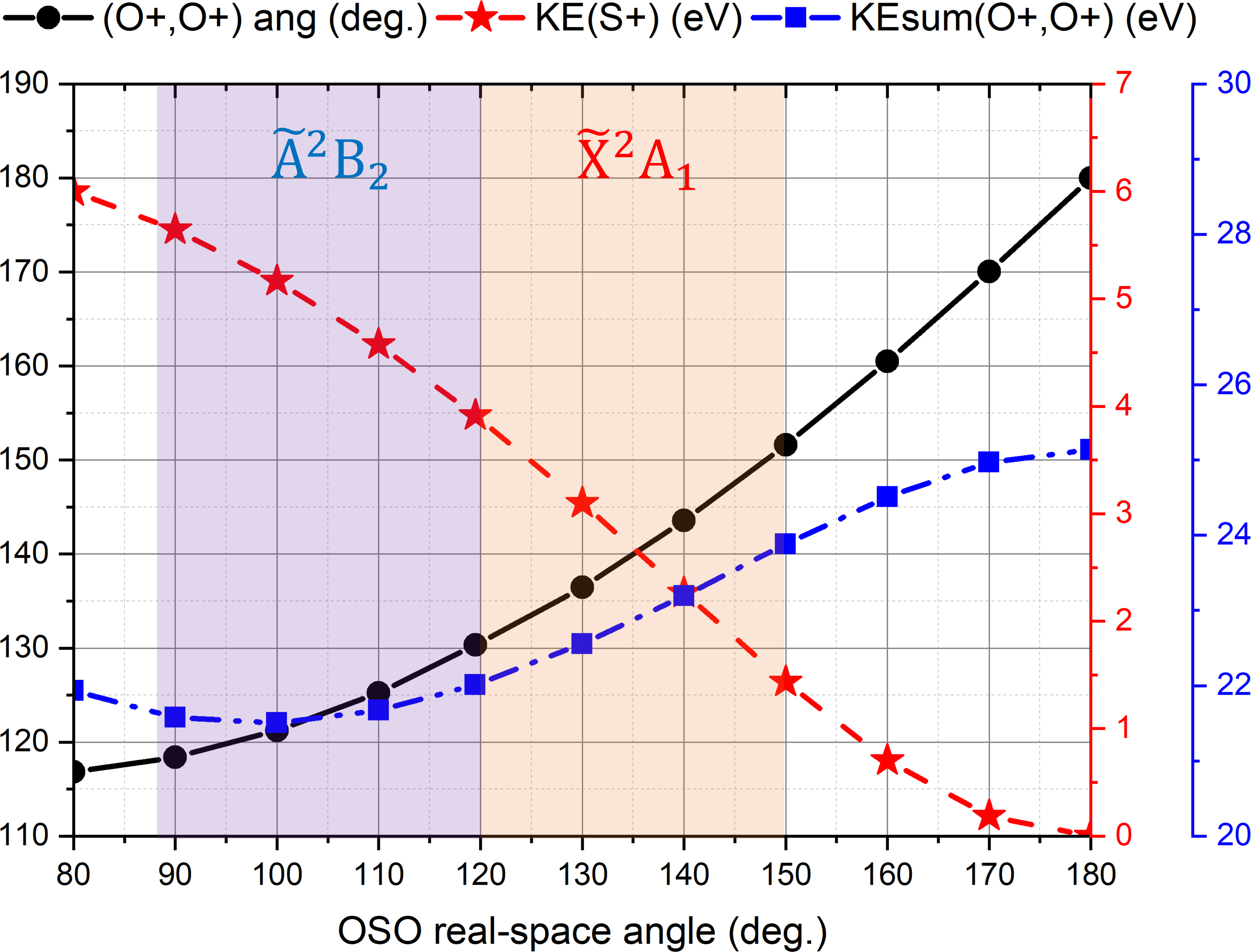}
  \caption{Simulation showing the dependence of the $\mathrm{\angle (O^+,O^+)}$, KE(S$^+$), and KEsum(O$^+$,O$^+$) as a function of the real-space OSO bond angle. The shaded area indicates the expected ranges of the OSO bond angle during the bending motion on the ground and the first excited states of SO$_2^+$ with classical boundaries following vertical ionization from the ground state of SO$_2$. In this region, there is a clear correlation between observables: KE($\mathrm{S^+}$) decreases while KEsum($\mathrm{O^+,O^+}$) and $\mathrm{\angle (O^+,O^+)}$ increases monotonically with larger OSO bond angle. The $\mathrm{\angle (O^+,O^+)}$ manifests the same behavior as the OSO bond angle of the nuclear wave packet propagating in the ionic ground state, giving an intuitive picture of the nuclear motion. In this simulation, the potential of the final charge state is assumed to be Coulombic potential, where each atom has a +1 point charge.}
  \label{fig:corr}
\end{figure}
\begin{figure}
\includegraphics[width=\columnwidth]{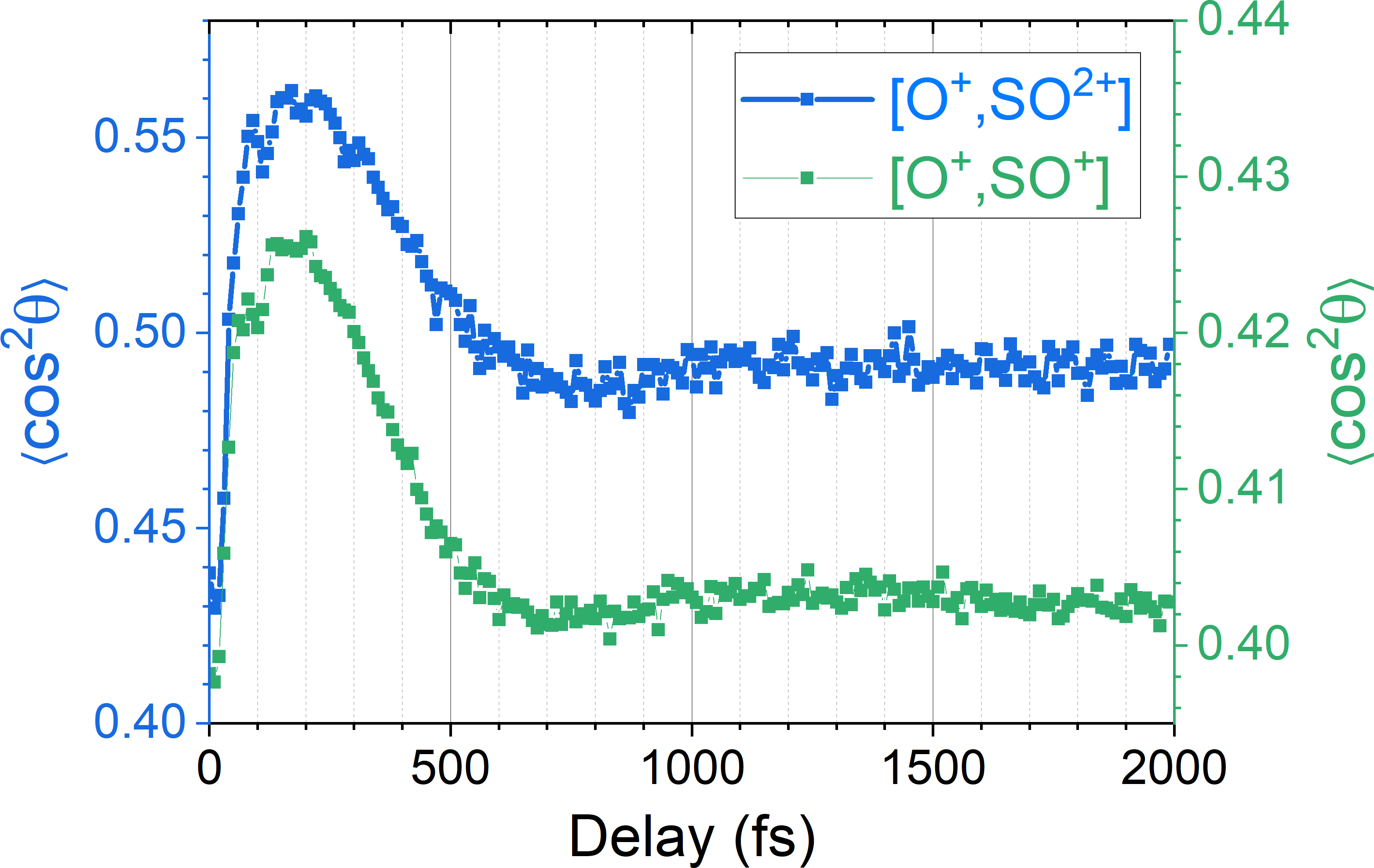}
  \caption{Delay dependence of $\langle \cos^2{\theta} \rangle$, where $\theta$ is the angle between the laser polarization and the recoil axis for two-body coincidence channels: (O$^+$, SO$^{2+}$) (blue) and (O$^+$, SO$^{+}$) (green). The extracted rotation time scale is consistent with the time scale deduced from the three-body (O$^+$,O$^+$,S$^+$) coincidence channel shown in the main paper.}
  \label{fig:rot_twobody}
\end{figure}
\begin{figure}
\includegraphics[width=\columnwidth]{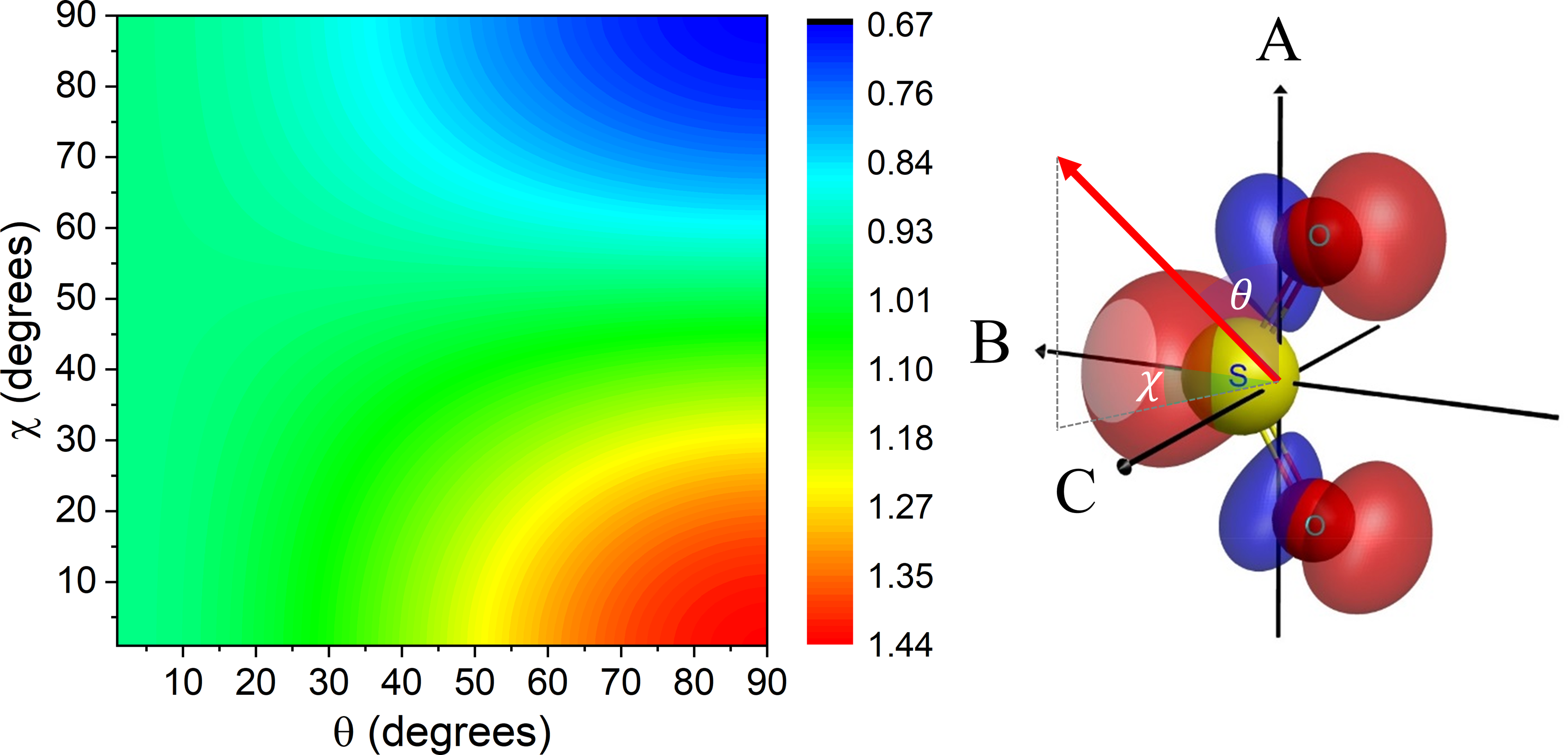}
  \caption{\textit{Left panel}: Experimentally determined angle-dependent strong-field ionization rate from SO$_2$ to SO$_2^+$ by a linearly polarized pulse (adapted from Ref.~\cite{Lam2021D}, Sec.~4.2). This result is derived from the delay-dependent yield of SO$_2^+$ in a pump-probe experiment, where the pump pulse aligns the neutral molecules and the probe pulse ionizes them, using the ORRCS algorithm developed in \cite{makhija2016, Lam2020, Wangjam2021, Lam2022}. \textit{Right panel}: Density profile of the highest occupied molecular orbital (HOMO). In the molecular frame, the laser polarization is described by the Euler angles $\theta$ and $\chi$, where $\theta$ is the polar angle (with the A axis), and $\chi$ is the azimuthal angle (on the BC plane). The ionization rate is highest along the B axis ($\theta=90^\circ,\chi=0^\circ$) and lowest along the C axis ($\theta=90^\circ,\chi=90^\circ$). The angles range from $0^\circ$ to $90^\circ$ because the molecules are aligned but not oriented. In other words, we cannot tell the difference between molecules that rotate $180^\circ$ about the principal axes.}
  \label{fig:ang_dep_ionization}
\end{figure}
\begin{figure}
\includegraphics[width=\columnwidth]{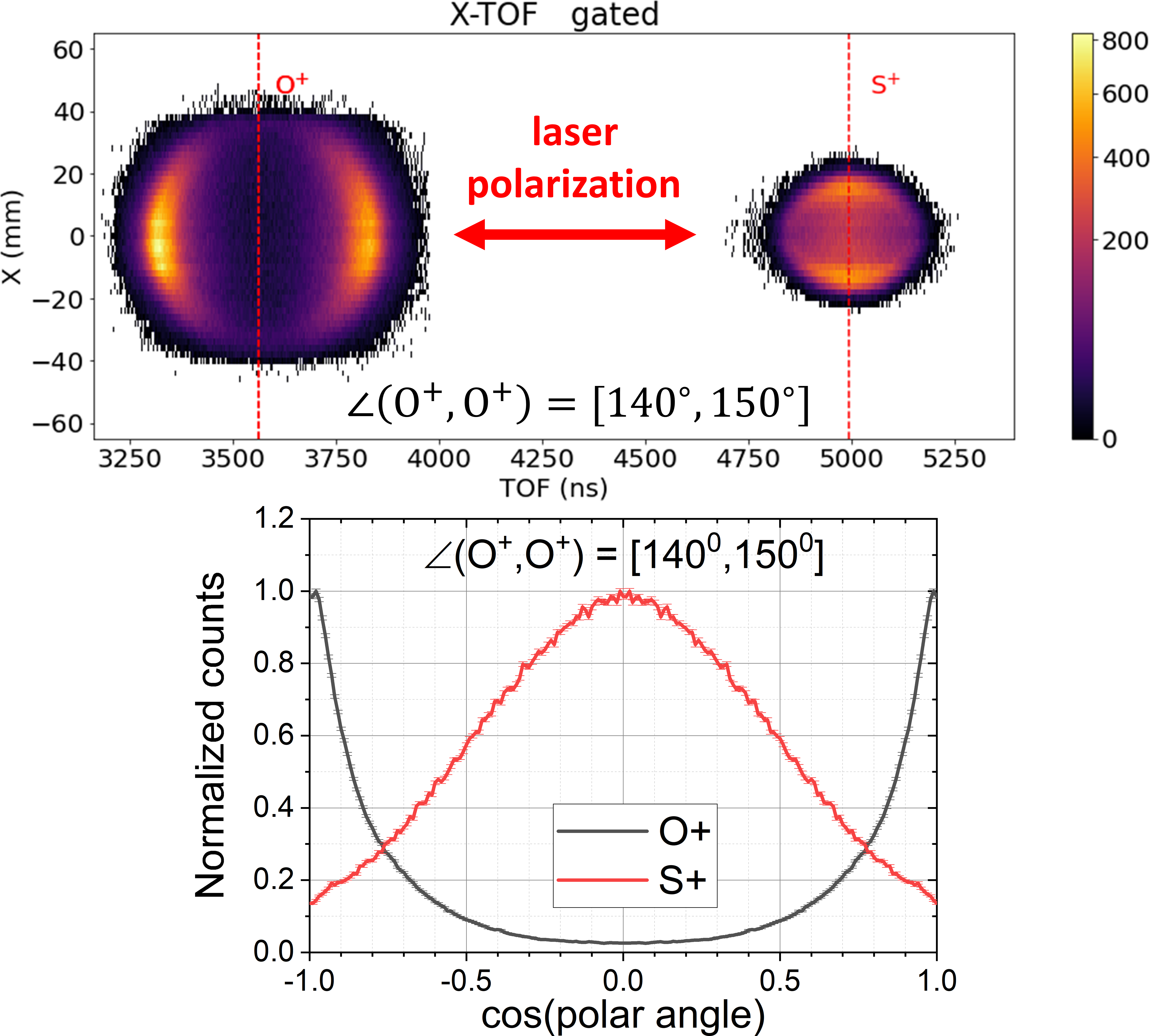}
  \caption{\textit{Top panel}: Position versus time-of-flight (TOF) spectrum. Only events with large $\mathrm{\angle (O^+,O^+)}$, from 140$^\circ$ to 150$^\circ$, were selected. Most of these events are excited by the pump to SO$_2^+$$({X}^2A_1)$ state before being fragmented into the triply charged channel by the probe. The O$^+$ fragments have denser distribution along the laser polarization axis, while the S$^+$ fragment distributes more perpendicular to this axis, suggesting that the probe preferentially ionizes aligned molecules to (O$^+$, O$^+$, S$^{+}$) channel. \textit{Bottom panel}: Polar angle distributions of O$^+$ and S$^+$ in the laboratory frame. On the $x$ axis, $0$ means perpendicular to the laser polarization, while $-1$ and $+1$ mean along the laser polarization.}
  \label{fig:pos_tof}
\end{figure}

List of figures and movies:
\begin{itemize}
    \item Fig.~\ref{fig:data}: Original data before subtraction of probe-only contribution.
    \item Fig.~\ref{fig:fft}: Comparison between the experimentally obtained and theoretically simulated FFT spectra of the as a function of $\angle$(O$^+$,O$^+$) angle.
    \item Fig.~\ref{fig:corr}: Simulation showing the dependence of the $\mathrm{\angle (O^+,O^+)}$, KE(S$^+$), and KEsum(O$^+$,O$^+$) as a function of the real-space OSO bond angle.
    \item Fig.~\ref{fig:rot_twobody}: Delay dependence of $\langle\cos^2{\theta}\rangle$ for two-body coincidence channels: (O$^+$, SO$^{2+}$) and (O$^+$, SO$^{+}$), where $\theta$ is the angle between the laser polarization and the recoil axis.
    \item Fig.~\ref{fig:ang_dep_ionization}: Angle-dependent strong-field ionization rate from SO$_2$ to SO$_2^+$ by a linearly polarized pulse obtained from an independent experiment.
    \item Fig.~\ref{fig:pos_tof} Position vs.~time-of-flight spectrum and the polar angle distributions of fragment momenta in the laboratory frame.
    \item Movie \textcolor{blue}{1}: A timed sequence of Newton plots visualizing the structural changes as a function of pump-probe delay from 0 to 390\,fs.
    \item Movie \textcolor{blue}{2}: A timed sequence of Newton plots visualizing the structural changes during one cycle of the bending vibration in the ionic ground state, repeated (from 170\,fs to 250\,fs).
    
\end{itemize}

\clearpage


\clearpage

\typeout{get arXiv to do 4 passes: Label(s) may have changed. Rerun}